\documentclass[%
reprint,aps,amsmath,amssymb,prl,superscriptaddress%
]{revtex4-2}
\usepackage{graphicx} 
\usepackage{dcolumn} 
\usepackage{bm} 
\usepackage[utf8]{inputenc}
\usepackage[T1]{fontenc}
\usepackage{etoolbox}
\usepackage{physics}
\usepackage{hyperref} 

\begin{document}

\title{Total Angular Momentum Conservation in {\em Ab Initio} Born-Oppenheimer Molecular Dynamics}
 
\author{Xuezhi Bian}
\affiliation{Department of Chemistry, University of Pennsylvania, Philadelphia, Pennsylvania 19104, USA}
\author{Zhen Tao}
\email{taozhen@sas.upenn.edu}
\affiliation{Department of Chemistry, University of Pennsylvania, Philadelphia, Pennsylvania 19104, USA}
\author{Yanze Wu}
\affiliation{Department of Chemistry, University of Pennsylvania, Philadelphia, Pennsylvania 19104, USA}
\author{Jonathan Rawlinson}
\affiliation{Department of Mathematics, University of Manchester, Manchester M13 9PL, UK}
\author{Robert G. Littlejohn}
\affiliation{Department of Physics, University of California, Berkeley, California 94720, USA}
\author{Joseph E. Subotnik}
\email{subotnik@sas.upenn.edu}
\affiliation{Department of Chemistry, University of Pennsylvania, Philadelphia, Pennsylvania 19104, USA}

\date{\today}

\begin{abstract}
    We prove both analytically and numerically that the total angular momentum of a molecular system  undergoing adiabatic Born-Oppenheimer dynamics is conserved only when pseudo-magnetic Berry forces are taken into account. This finding sheds light on the nature of Berry forces for molecular systems with spin-orbit coupling and highlights how {\em ab initio} Born-Oppenheimer molecular dynamics simulations can successfully capture the entanglement of spin and nuclear degrees of freedom as modulated by electronic interactions.
\end{abstract}

\maketitle 
 
\section{Introduction}
Born-Oppenheimer (BO) theory \cite{Born1927,Born1955} lies at the heart of two of the central problems in chemical physics -- electronic structure theory and molecular dynamics. Under the Born-Oppenheimer theory, the total molecular wavefunction is separated into two components according to a mass difference:  for the lighter electrons, one solves the electronic structure problems at different fixed nuclear geometries; for the slower nuclei, we simulate motion on a single (or sometimes on many coupled) electronic potential energy surface(s). As first pointed out by Mead and Truhlar in molecular systems \cite{Mead1979} and then derived by Berry more generally \cite{Berry1984}, such separation can lead to a nontrivial geometric phase and in general a gauge potential in the BO Hamiltonian \cite{Bohm2003}. 

Many intriguing phenomena have emerged from such a gauge structure. From the phase of an electronic eigenfunction transported around a conical intersection singularity, it is known \cite{Mead1992, Kendrick1997,Kendrick2003, Requist2016}  that a nuclear wavefunction will experience a sign-change during cyclic motion. This molecular geometric phase effect is a topological effect and has been observed in various experiments \cite{Schnieder1995,Yuan2018}. Pseudo-electromagnetic fields also arise from the gauge structure in Born-Oppenheimer theory, leading to non-trivial forces applied onto the nuclear wavefunction \cite{Berry1993,Kolodrubetz2017, Martinazzo2022}. For the most part, the pseudo-electric contribution (i.e. the diagonal Born-Oppenheimer correction) is thought to be of less importance than the  pseudo-magnetic (i.e. Berry force) contribution \cite{Tully2000}.    
Recently, experimental and theoretical developments in the solid-state have suggested that such pseudo-magnetic fields may lead to the phonon Hall effect \cite{Strohm2005,Zhang2010,Qin2012,Saito2019} and the phonon contribution in the Einstein-de Haas effect \cite{Zhang2014}.  In particular, the suggestion has been made that, in the presence of the pseudo-magnetic gauge field, phonon modes can carry a non-zero angular momentum \cite{Zhang2015,Saparov2022}.

Although the notion of a chiral phonon carrying angular momentum induced by interactions with electronic and spin degrees of freedom is becoming an active area of research nowadays \cite{Zhu2018,Chen2018}, molecular analogues of such physics are not as well known and the relationship between the pseudo-magnetic gauge field and nuclear angular momentum has not been fully explored. The most important contributions so far have come from Li and co-workers, who used an exact factorization approach \cite{Abedi2010} to study angular momentum transfer between electrons and nuclei \cite{Li2022}. 

To understand the problem in detail, let us decompose the total angular momentum into its nuclear, electronic orbital and electronic spin components (ignoring nuclear spin),
\begin{equation}
    {\bm J}_{\rm tot} = {\bm J}_{\rm nuc} + {\bm J}_{\rm orb} + {\bm J}_{\rm spin}.
\end{equation}
Due to the isotropy of space, the total angular momentum $\bm J_{\rm tot}$ must be conserved. However, neither $\bm J_{\rm nuc}$ nor $\bm J_{\rm ele}$ or $\bm J_{\rm spin}$ is a good quantum number; in principle, angular momentum can transfer between spins, electrons and nuclei in the presence of spin-electronic-rovibrational couplings. The magnitudes of the fluctuations of these observable quantities is of interest, especially insofar as the possibility that spin polarization may emerge from nuclear motion, which is one possibility for the chiral induced spin selectivity (CISS) effect \cite{Naaman2012,Naaman2019,Das2022}.  

%With this background in mind, here we investigate the direct impact of  the role of the pseudo-magnetic gauge field in the angular momentum transfer process.  
With this background in mind, in what follows, we consider a radical molecular system with spin-orbit coupling (SOC).  After reviewing the necessary equations of motion for propagating classical Born-Oppenheimer molecular dynamics (BOMD) with a pseudo-magnetic gauge field, we analytically calculate the change in angular momentum.  Note that, for a system with an odd number of electrons, electronic eigenstates always arise in Kramers degenerate pairs according to time-reversal symmetry. We will assume that  all dynamics follow one state in the Kramers degenerate set (which is equivalent to ignoring the off-diagonal component of the Berry curvature); this assumption represents an uncontrolled approximation, and yet running Hartree-Fock (HF) Born-Oppenheimer dynamics is fairly standard nowadays,  even in the case of a system with an odd number of electrons, given the cost of electronic structure calculations. \cite{Hammes1993,Culpitt2021}
%In other words, our conclusions below are applicable for molecular dynamics simulations where potential energy surfaces are generated by running time-reversal symmetry broken HF or DFT calculations (which is quite common nowadays) \cite{Neese2004,Noodleman1988}.  
Within this Born-Oppenheimer assumption, we show that (unlike the linear momentum) the total angular momentum is conserved only if we include the pseudo-magnetic gauge field that allows for angular momentum exchange between individual components.
As a proof of concept, we perform real-time {\em ab initio}  simulations  for methoxy radical isomerization, an open-shell molecular system with SOC, and we identify the magnitudes of the relevant fluctuations in each individual component.
We demonstrate that if we seek an {\em ab initio} framework for modeling spin dynamics in disordered thermal environments, pseudo-magnetic fields must be included within molecular dynamics simulations.

\section{Born-Oppenheimer molecular dynamics} 
For a general molecular Hamiltonian with spin-orbit coupling,
\begin{equation} 
\hat H = \hat T_{\rm n} + \hat H_{\rm el},
\end{equation} 
\begin{equation} \label{eq:Hel}
 \hat H_{\rm el} = \hat T_{\rm e} + \hat V_{\rm ee} + \hat V_{\rm en} + \hat V_{\rm nn} + \hat V_{\rm SO},
\end{equation} 
the total molecular wavefunction can be expressed as a sum of nuclear wavepackets multiplied by adiabatic electronic basis functions,
\begin{equation} \ket{\Psi({\bm R})} = \sum_j \chi_j({\bm R}) \ket{\Phi_j({\bm R})}.
\end{equation}

\noindent Here, the adiabatic electronic wavefunctions $\left\{\Phi_j({\bm r}, {\bm s}; {\bm R})\right\}$  are functions of electronic position and spin; they are constructed as solutions to the time-independent electronic Schr\"odinger equation at fixed nuclear coordinate $\bm R$: 
\begin{equation} \label{eq:HelS}
\hat H_{\rm el}({\bm R})\ket{\Phi_j({\bm R})} =  E_j ({\bm R})\ket{\Phi_j({\bm R})}.
\end{equation} 
According to the  Schr\"odinger equation, the nuclear wavepackets are propagated as
\begin{equation} \label{eq:nuclearH}
i\hbar\frac {\partial} {\partial t} \ket {\chi_j} = \sum_k H_{jk}^{\rm BO} \ket{\chi_k},
\end{equation}  
with the Born-Oppenheimer Hamiltonian:
\begin{equation} \label{eq:fullH}
\hat H_{jk}^{\rm BO} =  \sum_l \frac {({\hat {\bm P}} \delta_{jl} - {\bm A}_{jl}) \cdot (\hat {\bm P}\delta_{lk} - {\bm A}_{lk})} {2M} + E_j\delta_{jk}.
\end{equation} 
Here, $j,k$ label electronic eigenstates, ${\hat {\bm P}} = -i\hbar \bm \nabla$ is the nuclear momentum operator, ${\bm A}_{jk} = i\hbar\bra{\Phi_j} \bm \nabla \ket{\Phi_k}$ defines the Mead-Berry gauge potential.
The above equations are standard and all formally exact except one must truncate the electronic structure problem in Eq.~(\ref{eq:HelS}) to finite dimensions; for a brief discussion of the corrections that arise because of this truncation, see the Supplemental Material. 

% \tilde E_{jk} = E_j\delta_{jk}  + \frac 1 {2M}\left( \hbar^2 {\bra{\nabla\Phi_j}\ket{\nabla\Phi_k}} - \sum_l   {\bm A}_{jl}\cdot  {\bm A}_{lk} \right),

Unfortunately, propagating the coupled nuclear-electronic dynamics in Eq.~(\ref{eq:nuclearH}) is not possible for more than a few nuclear degrees of freedom (DoF) in practice because of the computational demands. Approximations must be made \cite{Coker1995,Curchod2018,Nelson2020}. One common approximation is to treat the nuclear DoFs classically and keep the electronic 
DoFs quantum mechanical which leads to a class of so-called "mixed-quantum-classical" methods \cite{Kapral1999,Crespo2018} (e.g., the mean-field Ehrenfest \cite{Ehrenfest1927,Meyera1979} and fewest switches surface hopping algorithms \cite{Tully1990,Wang2016,Subotnik2016}). An even simpler approach (that we will take in this letter) is to make the adiabatic approximation that the nuclei move slowly so that the system will stay on a single electronic eigenstate. In such a case, the effective Hamiltonian governing the system evolution simplifies to:
\begin{equation} \label{eq:HBO1}
\hat H^{\rm BO}_j =  \frac {({\hat {\bm P}} - {\bm A}_{jj})^2}  {2M} + E_j . 
\end{equation} 
If we define the nuclear kinetic momentum operator by $\hat {\bm \pi} = {\hat {\bm P}} - {\bm A}_{jj}$, note that one can derive the commutation relation,
\begin{equation} 
\left [ \hat \pi^{I\alpha}, \hat \pi^{J\beta} \right ] = i \hbar \Omega^{I\alpha J\beta}_{jj} ,
\end{equation} 
where the gauge-invariant Berry curvature tensor is defined as 
\begin{equation} 
\Omega_{jj}^{I\alpha J \beta} =  \left( \nabla_{I\alpha} A_{jj}^{J\beta} - \nabla_{J\beta} A_{jj}^{I\alpha} \right).
\end{equation}

For adiabatic BO dynamics, a simple application of the Heisenberg equations of motion yields the following:
\begin{equation}\label{eq:EOMR}
\frac {d  \hat R^{I\alpha}} {dt} = \frac i {\hbar} \left [ \hat H^{\rm BO}_j,  \hat R^{I\alpha} \right] = \frac {{\hat \pi^{I\alpha}}} {M^I},
\end{equation}
\begin{equation}
\begin{aligned} \label{eq:EOMpi}
\frac {d{\hat \pi}^{I\alpha}} {dt} &= \frac i {\hbar} \left [ \hat H^{\rm BO}_j, { \hat \pi^{I\alpha}} \right] \\
&= -\nabla_{I\alpha}  E_{j} +  \frac 1 2  \sum_{J,\beta}  \left( \Omega_{jj}^{I\alpha J\beta} \frac {   {\hat\pi}^{J\beta}}  {M^J} + \frac {   {\hat\pi}^{J\beta}} {M^J} \Omega_{jj}^{I\alpha J\beta} \right) .
\end{aligned}
\end{equation}
Here the index $I\alpha$ represents the Cartesian coordinate $\alpha = x,y,z$ of the $I$-th atom. According to Eq.~(\ref{eq:EOMpi}), the force on a nuclear wavepacket undergoing adiabatic motion can be separated into two parts: the first term is the usual BO force  and the second term is the pseudo-magnetic Berry force.  
At this point, if we also make the quantum-classical approximation, i.e., we
replace all nuclear operators by their classical variables, it is straightforward to propagate molecular dynamics on a single BO energy surface  provided we can compute the pseudo-magnetic Berry force,  
\begin{equation}
{\bm F}_{\rm Berry}^{I\alpha} =  \sum_{J,\beta}  \Omega_{jj}^{ I\alpha J\beta} \frac {   {\pi}^{J\beta}} {M^J} .
\end{equation} 
Before focusing on the question of angular momentum conservation within this BO framework, however, several computational and theoretical details must be addressed. 

First, the above equations of motion (Eqs.~(\ref{eq:EOMR})-(\ref{eq:EOMpi})) are valid only  in the adiabatic limit -- which usually requires that BO energy surface $E_j$ be well-separated from all other surfaces. Mathematically, one can rewrite the Berry curvature as follow:
\begin{equation} \label{eq:BC}
\begin{aligned}
    \Omega_{jj}^{ I\alpha J \beta } &= -\frac 2 \hbar {\rm Im} \sum_{k\neq j} A_{jk}^{I\alpha} A_{kj}^{J\beta}\\
    &=  -2\hbar {\rm Im} \sum_{k\neq j} \frac {\bra{\Phi_j} \nabla_{I\alpha} H_{\rm el} \ket{\Phi_k}\bra{\Phi_k} \nabla_{J\beta}H_{\rm el} \ket{\Phi_j}} {(E_j - E_k)^2}.
\end{aligned}
\end{equation}
According to Eq.~(\ref{eq:BC}), the Berry curvature is inversely proportional to the square of energy gap between the state of interest $\Phi_j$ and other states. From this expression, we may conclude that the pseudo-magnetic Berry force is essentially a nonadiabatic effect.  For the dynamics below, however, we will work in a regime where the nonadiabaticity is not overwhelming so that the adiabatic  approximation is not terrible (see below).

Second, we have not yet discussed the key issues of  electronic degeneracy or electronic structure, which are paramount when discussing Berry curvature for molecules. For a molecule with an odd number of electrons, Kramers' theorem ensures the doubly degeneracy of all the electronic eigenstates over the entire nuclear configuration space. Even for a single eigenenergy, Eqs.~(\ref{eq:HBO1}) - (\ref{eq:EOMpi}) are not enough, and instead a non-Abelian SU(2) gauge theory must be considered \cite{Mead1987,Koizumi1995}
\begin{equation} \label{eq:HBO2}
H^{\rm BO}_{j,\mu\nu} =  \sum_{\eta} \frac {({\bm P} \delta_{\mu\eta} -  {\bm A}_{j\mu j\eta}) \cdot ({\bm P} \delta_{\eta\nu}  - {\bm A}_{j\eta j\nu})}  {2M} +   E_{j}.
\end{equation} 
Here,  the two-fold degenerate eigenstates corresponding to the $j$-th eigenenergy are indexed by $\mu$ and $\nu$. The Hamiltonian described in Eq.~(\ref{eq:HBO2}) is gauge-covariant, i.e., the nuclear dynamics are independent of the choice of gauge of electronic eigenstates. 
Now, generating electronic states in the case of degeneracy is difficult.  
%In the context of trajectory-based dynamics, 
%single surface dynamics methods are valid only if the off-diagonal matrix elements of the Hamiltonian Eq.~\ref{eq:HBO2} are zero or small:
%\begin{equation} \label{eq:off0}
%    \bm P \cdot {\bm A}_{j\mu j\nu} \approx 0, \quad  \text{if} \quad \mu \neq \nu, 
%\end{equation}
%In such a case, the effective Hamiltonian is diagonal  (or nearly diagonal) and degenerate surfaces can be decoupled. Below, we will work with dynamics on such a surface with minimal gauge couplings.
%As mentioned above, we will ignore the impact of this degeneracy on the calculations below: we will effectively run on one electronic state  $\ket{\mu}$. See below.
For almost all {\em ab initio} dynamics calculations, one  approximates the electronic eigenstate by
a single Slater determinant (e.g., Hartree-Fock [HF] or density functional theory [DFT]) as a compromise between accuracy and computation time. Below, we too model dynamics with a generalized Hartree-Fock (GHF) \cite{Kubler1988} (effectively a non-collinear DFT \cite{Jimenez2011,Desmarais2019}) ansatz, where we include SOC. Of course, once the electronic structure packages generates one solution $\ket{\mu}$, we can also construct the corresponding  time reversed state $\ket{\nu} =  \hat T \ket{\mu}$ (for time reversible operator $ \hat T $). In principle, one might imagine that these two states themselves are not unique, as one can always rotate together any two degenerate solutions and find another solution. That being said, the linear combination of any two such states will no longer be a single Slater determinant; thus if one propagates along a GHF+SOC solution, the electronic structure method effectively chooses a gauge frame for the user.
Moreover, in the Supplemental Material, we show that, if we propagate the electronic wavefunction in the basis $\left\{\ket{\mu}, \ket{\nu} \right\}$ , more than 98\% of the resulting population still resides on state $\ket{\mu}$ after 120 fs and the fluctuations are weak. This empirical fact lends additional credence to our choice of running dynamics on one GHF state of a Kramer's pair (as mentioned above). 
% Luckily for us, for methoxy radical doublet dynamics (as studied below), the resulting GHF+SOC gauge is meaningful and satisfies the desired condition Eq.~(\ref{eq:off0}) approximately.  
%Therefore, for such a case, we can meaningfully study angular momentum conservation by following a given GHF+SOC trajectory. 
We will show below  (both analytically and computationally) that when we run dynamics along state $\ket{\mu}$, the total angular momentum is conserved if we include the pseudo-magnetic Berry force.

%for propagating BO molecular dynamics. In another word, the gauge selected by GHF+SOC method by enforcing the single Slater determinant constraint is dynamically meaningful. Therefore, we will use BO molecular dynamics with GHF+SOC method for a molecule with an odd number of electrons to demonstrate the interplay between pseudo-magnetic Berry force and angular momentum conservation.   

% is clearly difficult in general if we consider the simplest example of a complex-valued Hamiltonian involving a  that it can capture at least part of the nonadiabatic dynamical effect. Again, it is a limitation of BO representation, and one seeking for full description must go beyond single surface picture.  One simplest example is th

\section{Analytical Treatment of Angular Momentum Conservation} 
Within a BO representation for classical nuclei, the total linear momentum and angular momentum of a molecular system (moving along a single BO surface $E_j$ with electronic eigenbasis $\ket{\Phi_j}$) are given by the sum of the classical and quantum expectation values:
\begin{equation} \label{eq:Ptot}
{\bm P}_{\rm tot} = {\bm \pi}_{\rm n} + \left<{\hat {\bm P}}_{\rm e}\right> = {\bm \pi}_{\rm n} + \bra{\Phi_j({\bm R}(t))} {{\hat {\bm P}}_{\rm e}} \ket{\Phi_j({\bm R}(t))}, 
\end{equation}
\begin{equation}\label{eq:Jtot}
{\bm J}_{\rm tot} = {\bm J}_{\rm n}  + \left<{\hat {\bm J}}_{\rm e}\right> = {\bm R}_{\rm n} \times {\bm \pi}_{\rm n}  + \bra{\Phi_j({\bm R}(t))} {{\hat {\bm J}}_{\rm e}} \ket{\Phi_j({\bm R}(t))}.
\end{equation}
In Eq.~(\ref{eq:Ptot}), we have written ${\bm \pi}_{\rm n} = M \dot {\bm R}_{\rm n} $  to represent the kinetic (as opposed to canonical) momentum of the nuclear DoFs. 

%In classical BOMD, since the quantum nuclear wavefunction is simulated by classical trajectories, the canonical momentum is not measurable and replaced by the kinetic momentum, and electronic properties are measured at a given nuclear geometry. 

Now, BO theory requires wavefunction gauge conventions \cite{Littlejohn2023}, and for any electronic wavefunction, the only meaningful choice of gauge consistent with semiclassical theory satisfy translational and rotational invariance: 
\begin{equation} \label{eq:translation} 
    \left( \hat {\bm P}_{\rm e}  + \hat {\bm P}_{\rm n}  \right) \ket{\Phi_j({\bm R}(t))} = 0,
\end{equation}
\begin{equation}  \label{eq:rotation} 
    \left( \hat {\bm J}_{\rm n} + \hat {\bm J}_{\rm e } \right)\ket{\Phi_j({\bm R}(t))}  = 0,
\end{equation}
where $\hat {\bm J}_{\rm e} = \hat {\bm J}_{\rm orb}  + \hat {\bm J}_{\rm spin} $. 

From Eqs. \ref{eq:Ptot} and \ref{eq:translation}, we can express the total linear momentum as:
\begin{align} 
\label{eq:dPtot1}
    \sum_{I} \frac{d{P}^{I\alpha}_{\rm tot}}{dt} &=  \sum_{I} \left( \frac{d{\pi}^{I\alpha}_{\rm n}}{dt} - \frac{d\bra{\Phi_j} \hat {P}^{I\alpha}_{\rm n} \ket{\Phi_j}  } {dt} \right)\\
    \label{eq:dPtot2}
&=
\sum_{I}  \frac{d{\pi}^{I\alpha}_{\rm n}}{dt} + \frac {dA^{I\alpha}_{jj}} {dt} \\
\label{eq:dPtot3}
 & = 
 \sum_{I} \frac{d{\pi}^{I\alpha}_{\rm n}}{dt} + \frac{2}{\hbar}
 \sum_{I,J,k,\beta} \dot R_{\rm n}^{J\beta} {\rm Im} \left( A_{jk}^{I\alpha}A_{kj}^{J\beta} 
 \right) .
\end{align}
 
A similar expression can be derived for the total angular momentum using Eqs.~(\ref{eq:rotation}) and (\ref{eq:Jtot}).
\begin{equation} \label{eq:dJtot}
    \begin{aligned}
    \sum_{I} \frac{d{J}^{I\alpha}_{\rm tot}}{dt} &=  \sum_{I} \left( \frac{d{J}^{I\alpha}_{\rm n}}{dt} - \frac{d\bra{\Phi_j} \hat {J}^{I\alpha}_{\rm n} \ket{\Phi_j}  } {dt} \right) \\ 
    &=    \sum_{I} \frac{d{J}^{I\alpha}_{\rm n}}{dt} - \sum_{I, \beta,\gamma \neq \alpha}   \epsilon_{\alpha\beta\gamma}  R_{\rm n}^{I\beta} \frac{d\bra{\Phi_j} \hat {P}^{I\gamma}_{\rm n} \ket{\Phi_j}  } {dt}    \\ 
     &=    \sum_{I} \frac{d{J}^{I\alpha}_{\rm n}}{dt} + \frac 2 \hbar \sum_{\substack{I,J,k,\delta,\\ \beta ,\gamma \neq \alpha}}   \epsilon_{\alpha\beta\gamma}  R_{\rm n}^{I\beta}   \dot R_{\rm n}^{J\delta} {\rm Im}   
    \left( A_{jk}^{I\gamma}A_{kj}^{J\delta}  \right) . \\
    \end{aligned} 
\end{equation}

We can now easily demonstrate that the total linear and angular momentum will be conserved for BO trajectories if and only if  we use classical equations of motion that  include the Berry force
(and are consistent with Eqs.~(\ref{eq:EOMR}) and (\ref{eq:EOMpi})):
\begin{align}
\label{eq:EOMclassicalR}
\frac {d {\bm R}_{\rm n} } {dt} &= \frac {\bm \pi_n} {M}, \\
\label{eq:EOMclassicalP}
\frac {d {\bm \pi}_{\rm n} } {dt} &= -{\bm \nabla} E +  {\bm F}_{\rm Berry}.
\end{align}

Consider first the case of linear momentum: the right hand side of Eq.~\ref{eq:dPtot3} obviously vanishes if we plug in Eq.~\ref{eq:EOMclassicalP}. At the same time, note that, in practice, we usually  
include electronic translations factors (ETF) within  electronic structure calculations \cite{Fatehi2012,Bates1958,Delos1981,Illescas1998} so that $\sum_{I} A_{kj}^{I\alpha} = 0$.
Moreover, $\sum_I \dot {P}_{\rm n}^{I\alpha} = 0$ whenever the BO forces in Eq.~(\ref{eq:EOMclassicalP}) arise from a Hamiltonian with translational invariance (as follows from Noether's theorem).
Thus, there are effectively two ways to conserve linear momentum within a BO calculation:
either $(i)$ one can properly calculate the Berry curvature dressed with ETFs; or $(ii)$ one can work in a quick and dirty fashion, ignoring Berry curvature and equating canonical and kinetic momentum. In both cases, one will conclude that linear momentum is conserved.

Second, consider the case of angular momentum. Using Eqs.  \ref{eq:EOMclassicalR}, \ref{eq:EOMclassicalP} and the definition of $\bm J_{\rm n}$ in Eq.~\ref{eq:Jtot}, it follows that 
\begin{equation}
\frac {d {\bm J}_{\rm n} } {dt} =  {\bm R}_{\rm n} \times {\bm F}_{\rm Berry},
\end{equation}
and therefore the right hand side of  Eq.~(\ref{eq:dJtot}) also clearly vanishes. 
Note that, unlike the case of linear momentum,  angular momentum conservation can be achieved if and only if we include the Berry force in Eq.~(\ref{eq:EOMclassicalP}): the nuclear, electronic orbital and electronic spin exchange angular momenta back and forth
in a very complicated fashion that cannot be decomposed trivially as in the case of linear momentum.

 \section{Computational Treatment of Angular Momentum Conservation: Methoxy Radical Isomerization}
As an  example of the theory above, we have propagated BOMD on one of the ground doublet states of a methoxy radical. The potential energy surface is computed by GHF+SOC method with 6-31G(d,p) basis set and we have included the one-electron Breit-Pauli form of the  spin-orbit interaction. Details of the nuclear Berry force computation are given in Supplemental Material. The electronic structure was implemented in a local branch of Q-Chem 6.0 \cite{Epifanovsky2021}.  
\begin{figure} 
\includegraphics[width=0.5\textwidth]{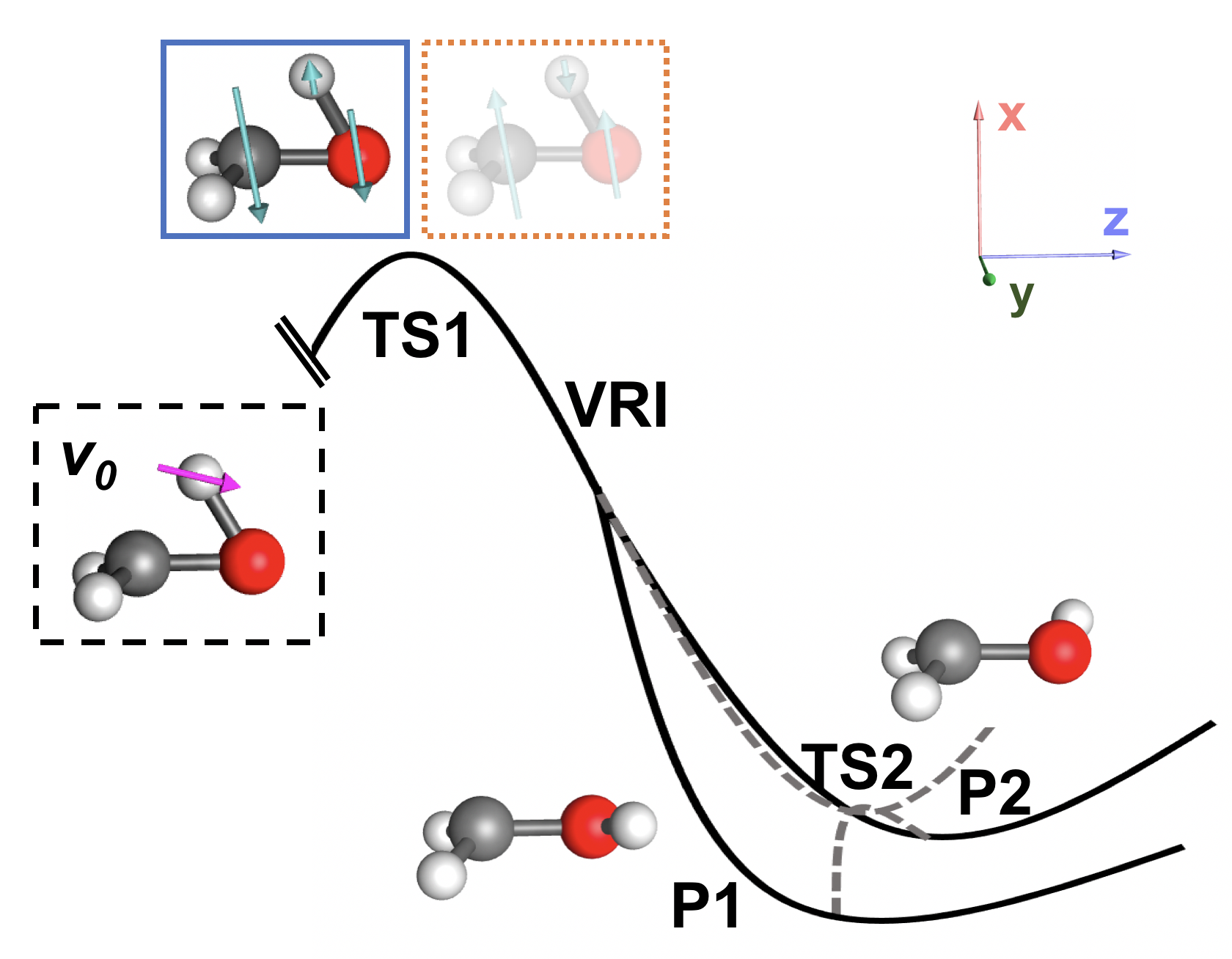} 
\caption{\label{fig:geometry} A schematic figure of methoxy radical isomerization reaction. The methoxy radical is initialized at TS1 with velocity $\mathbf{V}_0$ along the GHF-SOC solution with the majority spin down in the $x$-direction (as illustrated by the blue box above TS1). The magnitude of $V_0$ is 0.006 kcal/mol with a direction corresponds to the eigenvector with an imaginary eigenvalue of Hessian at TS1. This system is known to have a bifurcation (VRI point) in the reaction pathway and two reaction channels P1 and P2 are possible.} 
\end{figure}
 The methoxy radical isomerization is interesting because of the presence of a valley-ridge inflection point: the isomerization goes through one transition state (so-called TS1) and thereupon the reaction path bifurcates and allows for two different product wells (P1 and P2) \cite{Tao2022}. For the  post-transition state bifurcation simulations below, all methoxy radical dynamics were initiated at the transition state 1 (TS1) as computed via an  UHF calculations.\cite{Taketsugu1996} The trajectory was propagated with a step size of 2.5 a.u. (0.06 fs). The initial geometry and velocity are shown in Fig.~\ref{fig:geometry}.
 All dynamics are run along a single, smoothly varying GHF+SOC state (and there are, of course, two such states).

 %predicts one symmetry broken ground state and its time reversal pair with the opposite spin direction. For the simulations below, we have verified that the derivative coupling between these two states is less than $10^{-3}$ a.u., so that motion along one of these symmetry broken states is meaningful.

\begin{figure} 
\includegraphics[width=0.5\textwidth]{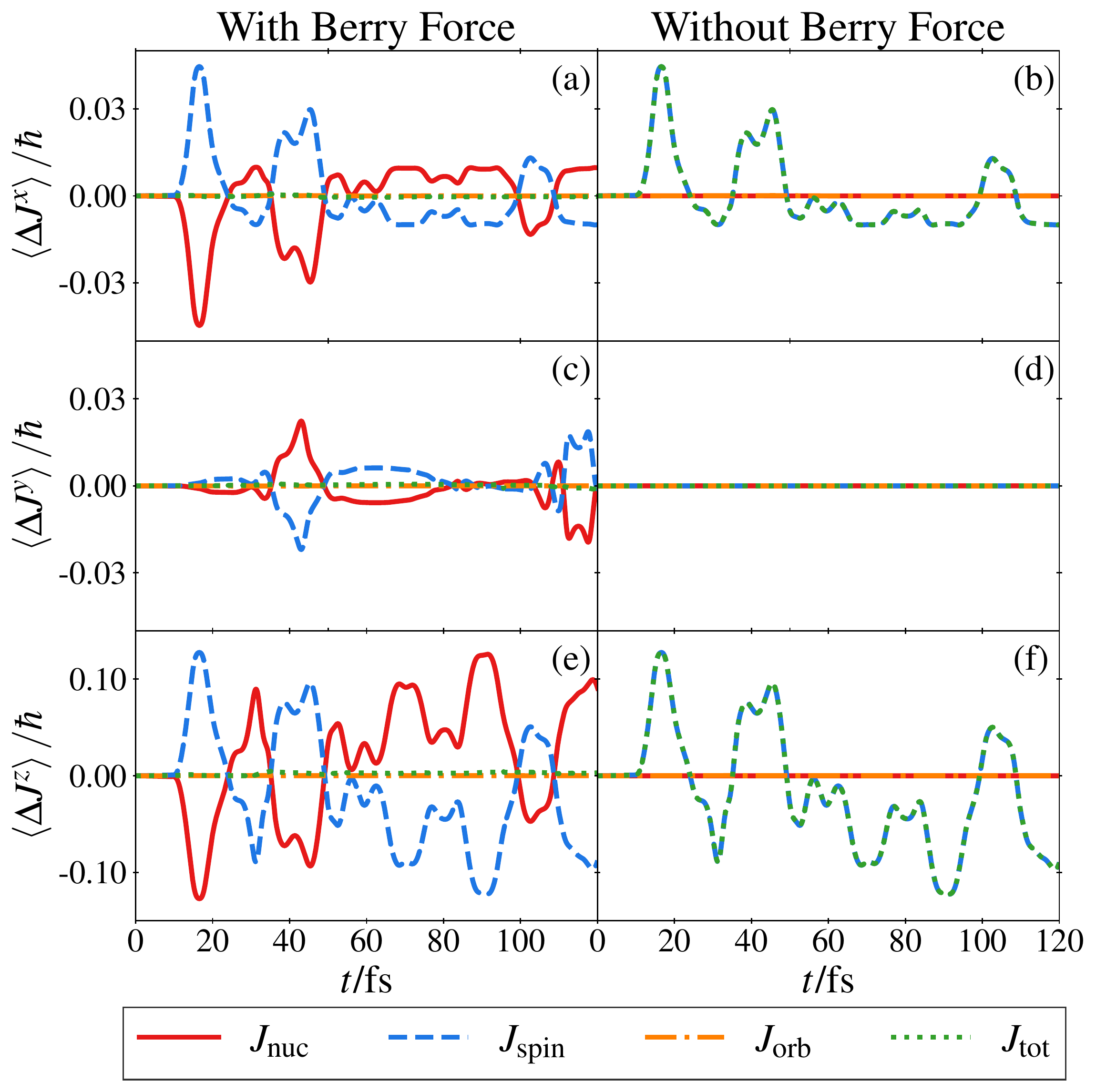} 
\caption{\label{fig:angmom}
The change in  the real-time  angular momentum $\left< \Delta J^\alpha (t)\right> = \left<  J^\alpha (t)\right>- \left< J^\alpha (0)\right>$ (relative to time zero) according to
a BOMD trajectory of methoxy radical isomerization reaction with Berry force and without Berry force.
For values at time zero, see Table \ref{tab:table1}.  Figs.(a),(c) and (e)  correspond to the Cartesian coordinates $x$, $y$ and $z$ shown in Fig.~\ref{fig:geometry} (and the same for Figs.(b),(d) and (f)).  
In Figs.(a),(c) and (e), with Berry force, the nuclear and spin angular momentum transfer from each other, resulting in a conservation of total angular momentum. By contrast, without Berry force [as shown in Fig.(b) and (f)], the total angular momentum is not conserved. The lack of a momentum change in the y-direction in Fig. (d) is due only to the symmetry of the potential and the choice of initial velocity.}
\end{figure}

In Fig.~\ref{fig:angmom}, we plot the  total angular momentum  change as a function of time relative to time zero, $\left<\Delta J^{\alpha}\right> = \left< J^{\alpha}(t)\right>- \left<J^{\alpha}(0)\right>$. The initial values for $J$ are reported in Table.~\ref{tab:table1}. We plot the  individual nuclear, electronic and spin components as calculated with (left hand side [a],[c],[e]) and without Berry force (right hand side [b],[d],[f]) as a function of time.   
As illustrated by Figs.~\ref{fig:angmom} (a)(c)(e), the total angular momentum is conserved when including the Berry force. The nuclear angular momentum is nearly equal and opposite to the spin angular momentum, while the electron orbital contribution is negligible.  In Figs.~\ref{fig:angmom}(b)(f), the results without Berry force show that the spin angular momentum changes are the same order of magnitude as in the case with Berry force, but the nuclear component is close to 0, so that there clearly is a violation of total angular momentum conservation. Note that in (d), there is no momentum change only because the nuclear motion is initialized from TS1 and there is symmetry along the y-axis. Adding the Berry force breaks this symmetry but still conserves the total angular momentum in (b). 
\begin{table}[b] 
\caption{\label{tab:table1}The initial values for the different components of the angular momentum for the trajectory recorded in Fig.~\ref{fig:angmom}. The classical nuclear and quantum electronic orbital angular momentum are both evaluated relative to the nuclear center of mass.}
\begin{ruledtabular}
\begin{tabular}{c c c c c} 
    & $J^{\alpha}_{\rm nuc}(0)/\hbar$  & $J^{\alpha}_{\rm spin}(0)/\hbar$ & $J^{\alpha}_{\rm orb}(0)/\hbar $   &  $J^{\alpha}_{\rm tot}(0)/\hbar$ \\
    \hline
    $\alpha= x$ &$0.00$ & $-0.49$&  $0.00$ & -$0.49$ \\
    $y$ & $0.00$ & $0.00$ & $0.00$& $0.00$\\
    $z$ & $0.00$ &$0.10$  & $0.00$& $0.10$
\end{tabular}
\end{ruledtabular}
\end{table}
 
According to Figs.~\ref{fig:angmom}(a)(c)(e) and (b)(f), when we include a Berry force, there is a clear transfer of angular momentum between electrons and spin and nuclei within the BO representation, while the total angular momentum is conserved.   That being said, by comparing dynamics with and without Berry force, one does notice
that the changes in spin angular momentum
are small between the two calculations. In other words, there does not seem to be a large feedback mechanism; changing the nuclear dynamics by introducing a  Berry force does not seem to induce large changes in spin for this particular set of dynamics.  Whether these conclusions will hold more generally, however, is a very open question.  For the present simulations, the spin-orbit coupling is small ($<10^{-4} \rm a.u.$),  we do not allow for nonadiabatic transitions, and there are few nuclear degrees of freedom. Previous work would suggest that with more reaction channels available, the presence of many electronic states with the possibilities of conical intersections \cite{Wu2021}, and larger spin-orbit coupling, the feedback between nuclear and spin dynamics may well be larger. For this reason, the Berry force has recently attracted attention as a possible factor in chiral induced spin selectivity experiments.

%are very similar. This suggests that the pseudo-magnetic field has a small effect on nuclear motion, which may not even be observable in molecular systems experimentally, as the change in spin angular momentum in doublet systems will not exceed $1 \hbar$ at last. However, this is not the case since in reality there could be multiple reaction channels on potential energy surfaces. For the presenting methoxy radical, there are two possible PTSB pathways and different initial spin states can lead to different final nuclear geometries.  
 
In conclusion, we have used {\em ab initio} BOMD to illustrate that conserving total angular momentum requires including the pseudo-magnetic Berry force. For a trajectory simulating methoxy radical isomerization,  under the assumption that we move along a single state of a Kramer's doublet pair, including Berry force maintains angular momentum conservation \footnote{For the present simulations, using the dynamics integrator in the Supplemental Material,the Berry curvature as calculated in the Supplemental Material, and a time step of 2.5 a.u., we can easily keep all errors in the total angular momentum below $10^{-3}\hbar$}
whereas without Berry force, the total angular momentum undergoes fluctuations up to $\approx 0.1 \hbar$.  
The present results (for a system with an odd number of electrons)  are relevant for systems with an even number of electrons as well.  In such a case, indeed the true electronic ground state wavefunction must be real-valued due to time reversal symmetry \cite{Mead1979-2}, and hence the on-diagonal Berry curvature (Eq.~(\ref{eq:BC})) must be zero. However, approximate electronic structure methods like GHF or non-collinear DFT with spin-orbit coupling often break time reversal symmetry and yield complex-valued electronic states with non-zero Berry curvatures \cite{Bistoni2021,Culpitt2022}; and recent work has demonstrated that, for such cases, the non-zero Berry curvature (as calculated by an approximate mean-field Schr\"odinger equation) can be meaningful \cite{Bian2022-2}. %Nevertheless, when SOC is present and we can expect a complex-valued mean-field solution, working with systems with an even number of electrons can be very difficult; e.g. even a singlet-triplet intersystem crossing model should require four electronic states \cite{Bian2022-3}. For this reason, below, we focus on the simpler case of an odd number of electrons. 
Thus, in general, simulations of coupled spin-nuclear dynamics (with even or odd numbers of electrons) must include angular momentum correctly if one seeks either to model spin-lattice relaxation \cite{Torchia1982} and/or design and control spin-dependent chemical reactions with the spin DoF \cite{Steiner1989}.

Looking forward, one would very much like to find tractable trajectory methods that go beyond the Born-Oppenheimer approximation studied here\cite{Bian2021} and that can be used to check the validity of the present single-state simulations. In an upcoming publication, we will discuss the possibility of using Ehrenfest dynamics as one dynamical tool that conserves linear and angular momentum and incorporates different nonadiabatic effects (but does not include branching \cite{Tully1998}). 
Another interesting direction is phase-space surface-hopping \cite{Wu2022,Bian2022} which should allow for branching.
More generally, new, rigorous quantum and semiclassical nonadiabatic frameworks will be crucial in the future as far as going beyond the single-state Born-Oppenheimer approximation presented here and allowing us to disentangle how angular momentum is spread across multiple nuclear degrees of freedom and multiple electronic states (rather than just requiring that all motion occur along a single electronic state as we enforced here). 
The implications of Berry forces and angular momentum conservation as far modeling chiral induced spin separation \cite{Teh2022} and avian bird magnetic field reception \cite{Rodgers2009} remains a very exciting area of research at the intersection of spintronics, magnetochemistry, molecular dynamics, and light-matter physics for future development.

%Future performing {\em ab initio} nonadiabatic molecular dynamics that fully considers the gauge potential holds great promise and will provide deeper insights into the field of spin chemistry. 
 
\bibliography{main.bib} 
\end{document}